\documentstyle[preprint,aps]{revtex}

\newfont{\bg}{cmr10 scaled\magstep4}

\newcommand{\bigzerou}{\smash{\lower1.ex\hbox{\bg 0}}}

\newcommand{\AmS}{{\protect\the\textfont2
  A\kern-.1667em\lower.5ex\hbox{M}\kern-.125emS}}

\begin{document}
\draft
\preprint{KANAZAWA 95-11,hep-ph 9509266 (revised)}
\title{
Quark Confinement in QCD and New Bosons
}

\author{Tsuneo Suzuki
\footnote{ E-mail address:suzuki@hep.s.kanazawa-u.ac.jp }
}
\address{Department of Physics, Kanazawa University, Kanazawa 920-11, Japan}

\maketitle

\begin{abstract}
If the dual Meissner effect due to abelian monopole condensation
is the quark confinement mechanism of QCD 
as suggested in recent Monte-Carlo simulations of lattice QCD,
new axial-vector and scalar bosons with the mass of O(1GeV)
would appear as  physical 
states which are different from ordinary hadrons and glueballs.
The axial-vector boson can not decay into ordinary color-singlet hadrons 
and glueballs owing to a remainig global 
discrete permutation symmetry with respect to colors
 (Weyl symmetry) if the vacuum respects the symmetry as suggested from 
lattice MC simulations. 
\end{abstract}

\input epsf


\newpage
\narrowtext

\section{Introduction}
The 'tHooft idea\cite{thooft2} on quark confinement mechanism in QCD 
starts with partially gauge-fixing  color $SU(3)$ 
 in such a way that the maximal 
torus group $U(1)\times U(1)$ remains unbroken. 
This is called abelian projection.
After the abelian projection, monopoles appear and then 
QCD can be regarded as abelian $U(1)\times U(1)$ 
theory with electric charges (quarks and gluons) and magnetic charges 
(monopoles). 'tHooft conjectured, if the monopoles make condensation, 
electric charges and then quarks are confined due to a mechanism dual to 
the Meissner effect. 

Suppose the 'tHooft confinement mechanism is actually realized in QCD. Then 
after abelian projection, abelian components of gluons and abelian 
monopoles are expected to be 
essential dynamical quantities governing  quark confinement mechanism.
Numerical studies  have been done by many groups 
in order to test the confinement mechanism 
in the framework of lattice QCD. The present results are summarized as follows:
\begin{enumerate}
\item (Abelian dominance) Essential features of confinement such as the string 
tension seem to be explained in terms of $U(1)\ (\times U(1))$ 
operators composed of abelian link fields alone 
\cite{suzu90,hio91b,suzu93,suzu95a,miya95a,miya95b,suzu95c,ejiri95b,suzurev}
in the maximally abelian (MA) gauge\cite{kron,hio91a} 
and in some cases also in the Polyakov gauge.
\item (Monopole dominance) The $U(1)\ (\times U(1))$ 
operators are written by a 
product of two parts, a monopole current or Dirac string part 
and a photon part. The confinement phenomena seem to be reproduced well by 
the monopole part alone\cite{shier,hio91b,suzu93,kita93,stack,shiba94b,ejiri95a,suzu95a,miya95a,miya95b,suzu95c,ejiri95b,ejiri95c}. 
\item (Scaling of monopole density and monopole dynamics)
Monopoles in QCD seem to remain important in the continuum limit as seen from 
scaling behaviors\cite{born,hio91b,hio92,ivanenko,shiba94a,suzu95b,suzurev}.
Monopoles are jammed\cite{kron} and make a long connected loop 
in the confinement phase. The long loop seems to be responsible for 
confinement\cite{ejiri95a,shiba94a,kita95}. 
\item (The dual Meissner effect and flux squeezing)
Abelian electric (color) flux is seen to be squeezed 
and the QCD vacuum seems to 
be near the border between type 1 and type 2 magnetic superconductor
\cite{hay,cea,matsu94}. 
\item (Order parameters of confinement)
It is found a candidate of the order parameter of confinement which transforms 
under the dual $U(1)\ (\times U(1))$ symmetry nontrivially and which vanishes 
in the confinement phase\cite{giacomo}.
\item (Monopole action and monopole condensation)
The effective monopole action can be derived in $SU(2)$ and $SU(3)$ QCD.
The block-spin transformation on the dual lattice 
strongly suggests that $SU(2)$ QCD is always in the monopole condensed phase 
(and so in the confinement phase) for all $\beta$ in the infinite volume 
limit\cite{shiba94a,suzu95b,suzurev}.
\item (Gauge (in)dependence)
Gauge independence of the mechanism is the biggest problem to be proved.
\end{enumerate}

It is the aim of this note 
to show that, if the 'tHooft idea is realized in nature, there must appear 
new axial-vector and scalar bosons with the mass of O(1GeV) as 
physical states which are not confined. The field 
operators of the bosons  are not invariant
 under a global discrete Weyl 
transformation except their special combinations.  
If the vacuum also respects the Weyl symmetry,  
the states which are  globally color non-singlet are predicted to exist.
The new states are seen 
to have characteristic decay modes
into ordinary hadrons and glueballs 
which are trivial under the Weyl transformation. Experimental tests of 
such bosons are essential to prove the 'tHooft mechanism 
in addition to lattice Monte-Carlo simulations.

\section{Abelian projection}
Abelian projection of QCD is a partial gauge fixing leaving 
the maximal torus group unbroken. For example, it is 
done as follows.
Choose an operator $X(x)$ which transforms non-trivially 
under $SU(3)$ transformation:
\begin{eqnarray}
   A_{\mu 0} (x) & \to & A_\mu (x)= V(x)A_{\mu 0} (x)V^\dagger (x)
   - \frac{i}{g}\partial_\mu V(x)V^\dagger (x)\\
   \psi_0 (x)& \to & \psi (x)= V(x)\psi_0 (x).
\end{eqnarray}
Then abelian projection is to choose $V(x)$ so that $X(x)$ is diagonalized: 
\begin{eqnarray}
   X(x) \to \widetilde X(x) = diagonal. \label{wx}
\end{eqnarray}
It is known that, once an ordering of the diagonal elements of 
$\widetilde X(x)$ is chosen, the nonabelian part of the gauge
is fixed uniquely\cite{kron}. 
The diagonal element $d(x)$ of $SU(3)$ is not fixed.
$\{d(x)\}$ is the maximum torus group of $SU(3)$, which is the residual 
$U(1)\times U(1)$ gauge symmetry. 

Let us look at QCD at this stage without further fixing the gauge 
of the residual symmetry. 
First, we explore how the fields after the abelian projection 
transform 
{\it under an arbitrary $SU(3)$ gauge transformation $S(x)$.}
 Since $V(x)$ is a functional of  (gauge) fields and so it 
transforms non-trivially under $S(x)$.
Let us fix the form of $V(x)$ such that all diagonal components 
of the exponent of $V(x)$ are zero. This is always possible if one uses the 
residual symmetry. Then 
$V(x)$ is found to transform 
{\it under $S(x)$}
 as
\begin{eqnarray}
   V(x)\stackrel{S}{\longrightarrow}V^S (x)= 
d^S (x)V(x)S^\dagger (x). \label{vs}
\end{eqnarray}
$V^S (x)$ diagonalizes an operator which is transformed from 
$X(x)$ under $S(x)$.
$d^S (x)$ is necessary for $V^S (x)$ to take the same form as $V(x)$
fixing the arbitrariness due to the remaining $U(1)\times U(1)$.

The gauge field after the abelian projection, 
$A_\mu (x)$, transforms 
{\it under $S(x)$} as
\begin{eqnarray}
   A_\mu (x)\stackrel{S}{\longrightarrow} A_\mu^S (x) = 
d^S (x) A_\mu (x)d^{S\dagger}(x)
   - \frac{i}{g}\partial_\mu d^S (x)d^{S\dagger}(x). \label{atran}
\end{eqnarray}
After the abelian projection, $ A_\mu (x)$  
transforms only under the diagonal matrix $d^S (x)$.
Since the last term of (\ref{atran}) is composed of the diagonal part alone, 
the diagonal part of $ A_\mu (x)$ transforms like a photon.
The off diagonal part of $ A_\mu (x)$ transforms 
like a charged matter.
The quark field transforms {\it under $S(x)$} as
\begin{eqnarray}
   \psi (x)\stackrel{S}{\longrightarrow}\psi^S (x)
= d^S (x)\psi (x).
\end{eqnarray}
It is important that, after abelian projection,
 $\bar{\psi}^i (x)\psi^i (x)$ and 
$\psi^1 (x)\psi^2 (x)\psi^3 (x)$ 
are neutral and 
at the same time  invariant under any $SU(3)$ transformation $S(x)$.

The most interesting fact of abelian projection is that monopoles appear 
in the residual abelian channel.
We treat $SU(2)$ QCD for simplicity. 
After abelian projection, we define an abelian field strength as 
\begin{eqnarray}
   f_{\mu\nu}(x) = \partial_\mu A_\nu^3 (x) 
              - \partial_\nu A_\mu^3 (x).
\end{eqnarray}
$f_{\mu\nu}(x)$ can be rewritten 
in terms of the original field as 
\begin{eqnarray}
   f_{\mu\nu} (x)= \partial_\mu (\hat Y^a (x)A_{\nu 0}^a (x))
-\partial_\nu (\hat Y^a (x)A_{\mu 0}^a (x))
              - \frac{1}{g}\varepsilon_{abc}\hat Y^a (x)
\partial_\mu \hat Y^b (x)\partial_\nu \hat Y^c (x)
\end{eqnarray}
where $\hat Y(x) =V^{\dagger}(x) \sigma_3 V(x)=\hat Y^a (x)\sigma^a$.
$\hat Y^a (x)$ obeys 
\begin{equation}
\hat Y^a (x)\hat Y^a (x)=1.\label{ysphere}
\end{equation}
A current
\begin{eqnarray}
   k_{\mu}(x) &=& \frac{1}{2}\varepsilon_{\mu\nu\rho\sigma}\partial^\nu
             f^{\rho\sigma}(x)         \\
         &=& \frac{1}{2g}\varepsilon_{\mu\nu\rho\sigma}
             \varepsilon_{abc}\partial^\nu\hat Y^a (x)
\partial^\rho \hat Y^b (x)\partial^\sigma \hat Y^c (x)
\end{eqnarray}
is always zero if $V(x)$ is fixed. However,
at a point $x$ where the eigenvalue of the diagonalized operator 
$X(x)$ is degenerate, $V(x)$ is not well defined and $k_{\mu}(x)$ does 
not vanish there.
We calculate the charge in the three dimensional volume 
$\Omega$ around $x$:\cite{arafune}
\begin{eqnarray}
   g_m=\int_\Omega k_0 (x)d^3x
     &=& \frac{1}{2g}\int_\Omega\varepsilon_{0\nu\rho\sigma}
         \varepsilon_{abc}\partial^\nu\hat Y^a (x)\partial^\rho \hat Y^b (x)
         \partial^\sigma \hat Y^c (x)d^3x \\
     &=& \frac{1}{2g}\int_{\partial\Omega}\varepsilon_{ijk}
         \varepsilon_{abc}\hat Y^a (x)\partial_j \hat Y^b (x)
         \partial_k \hat Y^c (x)d^2\sigma_i \\
     &=& \frac{4\pi n}{g}  ,
\end{eqnarray}
where $n$ is an integer.
$n$ is a topological number corresponding to a mapping between 
the sphere (\ref{ysphere}) in the parameter space and the sphere 
$\partial\Omega$ of $\Omega$.
Because this equation represents the Dirac quantization condition, 
$g_m$ can be interpreted as a magnetic charge.
The monopole current $k_\mu (x)$ is a topologically conserved current
  $ \partial_\mu k^\mu (x) = 0 .$
{\it Abelian projected QCD can be regarded as 
an abelian theory with electric charges and monopoles.}
'tHooft\cite{thooft2} conjectured
if the monopoles condense, abelian charges are confined 
due to the dual Meissner effect. This means quark confinement.

\section{The Weyl symmetry }
Once an abelian projection is done with a choice of a certain 
gauge-fixing matrix,
 abelian charge neutrals are invariant 
also under color $SU(3)$ as proved above. However, since such a proof is 
done on a fixed gauge orbit, it 
 does not mean always that all abelian 
neutrals are also $SU(3)$ color singlets literally. 

Let us start with the usual $SU(3)$ QCD Lagrangian after abelian projection:
\begin{eqnarray}
L&=& \frac{1}{4}F_{\mu\nu}^{(a)}F^{(a)\mu\nu} + 
i\sum_{f}\bar{\psi}_f^i\gamma^{\mu}(D_{\mu})_{ij}\psi_f^j \nonumber \\
&& - \sum_{f}m_f\bar{\psi}_f^i\psi_f^i + L_{GF+FP}\ , \label{lag}
\end{eqnarray}
where 
\begin{eqnarray*}
F_{\mu\nu}^{(a)}&=&\partial_{\mu}A_{\nu}^a - \partial_{\nu}A_{\mu}^a
+ gf_{abc}A_{\mu}^b A_{\nu}^c\ ,\\
(D_{\mu})_{ij}&=& \delta_{ij}\partial_{\mu} 
- ig\sum_{a}\frac{\lambda_{ij}^a}{2}A_{\mu}^a
\end{eqnarray*}
and $L_{GF+FP}$ is a gauge-fixing term. For example, in the MA gauge,
\begin{eqnarray}
L_{GF+FP}=\delta_B\{\sum_{i\neq j} \bar{c}^{ji}
(\partial_{\mu} + ig(A_{\mu}^{ii}-A_{\mu}^{jj}))A^{\mu ij}\}, \label{mgf}
\end{eqnarray}
where $\delta_B$ is the BRS transformation,  
$\bar{c}$ is the Faddev-Popov ghost and
the gluon field $3\times 3$ matrix is 
\begin{equation}
A=\left( A^{ij} \right)
=\sum_{\alpha=1}^8 \frac{1}{2}A^{\alpha}\lambda_{\alpha}\ ,
\end{equation}
with  the GellMann matrices $\lambda_{\alpha}$.
In a unitary gauge where an adjoint operator $X$ is diagonalized,
\begin{eqnarray}
L_{GF+FP}=\delta_B\{\sum_{i\neq j} \bar{c}^{ji}X^{ij}\}. \label{ugf}
\end{eqnarray}
Note that one has to further 
fix the gauge of the remaining $U(1)\times U(1)$ in the continuum to get the 
Fadeev-Popov determinant.
Also the monopole contribution to the functional measure should be taken into 
account.

What symmetries are left unbroken after abelian projection? 
It is well known that maximally abelian torus group $U(1)\times U(1)$ 
is unbroken as a local symmetry. In addition, 
any global discrete permutation with respect to 
three colors makes the Lagrangian 
(\ref{lag}) and (\ref{mgf}) or 
(\ref{ugf}) unchanged. The discrete permutations compose the 
permutation group which is the Weyl group of $SU(3)$. 
The discrete symmetry corresponds to the fact that one can choose 
any ordering of the diagonal elements of 
$\widetilde X(x)$ in (\ref{wx}) in the case discussed above.
In $SU(3)$, one can choose any one of 
six $SU(3)/U(1)^2$ gauge-fixing matrices 
corresponding to six different orderings of three eigenvalues.
The Weyl transformation interchanges the different gauge-fixing matrices.
Hence the Weyl group is a subgroup of the original $SU(3)$.
Monopole physics are unchanged in any choice, since space-time points 
where the two eigenvalues are degenerate are the same and the topology is 
unchanged.

Consider for example a permutation $(12)$ which is, in the matrix 
representation, expressed by 
\begin{equation}
V_3=\left( \begin{array}{ccc}
0&-1&0\\
1&0&0\\
0&0&1
\end{array}
\right)
\in \ SU(3).
\end{equation}
From the transformation properties 
\begin{eqnarray}
\psi &\rightarrow& V_3 \psi\ ,\\
A &\rightarrow& V_3AV^{\dagger}_3\ ,
\end{eqnarray}
we get 
\begin{eqnarray}
\bar{\psi}\lambda_3 \psi &\rightarrow& -\bar{\psi}\lambda_3 \psi\ ,\\
\bar{\psi}\lambda_8 \psi &\rightarrow& \bar{\psi}\lambda_8 \psi\ ,\\
A^3 &\rightarrow& -A^3\ , \ \ A^8 \rightarrow A^8\ ,\\
A^{12}&\rightarrow& A^{21}\ ,\ \ A^{13}\rightarrow A^{23}\ ,\ \ etc.,
\end{eqnarray}
Similary, under a permutation $(31)$,  $A^3 + \sqrt{3}A^8$ and 
$\bar{\psi}(\lambda_3+\sqrt{3}\lambda_8)\psi$ change their signs.

Considering also the $U(1)\times U(1)$ property, one can see, 
for example, $\bar{\psi}\lambda_i \psi$ ($i=3,8$) 
and the phisical parts of $A^i$ ($i=3,8$) are $U(1)\times U(1)$ 
neutral but not invariant under the Weyl transformation. Namely 
these are not global $SU(3)$ color-singlets. However, since 
\begin{equation}
\sum_{i=1}^{3}\bar{\psi}_{0}^i\psi_{0}^i = 
\sum_{i=1}^{3}\bar{\psi}^{i}\psi^{i}\ ,
\end{equation}
$\bar{\psi}\psi$ is  global color-singlet. 

Such a state exists also in the case of baryons.
There are six $U(1)\times U(1)$ neutral  baryons 
$\psi_{f1}^1\psi_{f2}^2\psi_{f3}^3$ where $fi$ denotes the set of 
quantum numbers like flavor other than color.
It is possible to prove 
\begin{equation}
\sum_{i,j,k=1}^{3}\epsilon_{ijk}\psi_{0f1}^i\psi_{0f2}^j\psi_{0f3}^k = 
\sum_{i,j,k=1}^{3}\epsilon_{ijk}\psi_{f1}^i\psi_{f2}^j\psi_{f3}^k\ .
\end{equation}
Hence the antisymmetric combination is equal to the original color 
singlet baryon. However, 
other five combinations are $U(1)\times U(1)$ neutral, but Weyl variant. 

Existence of the remaining Weyl symmetry in generic abelian projection 
is proved as follows. Note that
one can always find an adjoint operator which is diagonalized 
under any abelian projection whereever 
the gauge-fixing matrix is well-defined. 
Define a gauge-fixing $SU(3)$ matrix $V(x)$
of an abelian projection. Then 
$\hat Y(x) =V^{\dagger}(x) \lambda_0 V(x)$ where $\lambda_0$ is any 
linear combination of $\lambda_3$ and $\lambda_8$ is a functional of 
gluon (quark) fields and transforms 
like an adjoint operator as seen from (\ref{vs}). 
An abelian projection can be 
characterized  as the diagonalization of the matrix $\hat Y(x)$, i.e.,
$\hat Y^{i\neq j}=0$ at any space-time point where monopoles do not exist.
These conditions are trivially Weyl symmetric. 
Monopoles exist where $V(x)$ and $\hat Y(x)$ are ill-defined. 
A Weyl transformation changes any $\hat Y(x)$ among the set of 
the matrices $V^{\dagger}(x) \lambda_0 V(x)$. Hence monopole physics 
remain unchanged.

\section{Dual Meissner effect as the dual Higgs mechanism}
The monopole condensation causes the dual Meissner effect and the quark 
confinement. The effect is a kind of the Higgs mechanism just 
as the usual Meissner effect in superconductivity. 
Here the theory is well described in terms of the dual variables after a dual 
transformation. The spontaneously broken symmetry is  
magnetic $U(1)\times U(1)$ 
which is dual to the remaining electric $U(1)\times U(1)$ maximal torus group
\cite{thooft2,suzu88,maedan89,maedan90}.
Hence axial vector massive gauge bosons and scalar (dual Higgs) bosons are 
predicted to exist just as in the usual Higgs mechanism.

The situations can be seen more clearly when one constructs 
a dual abelian $U(1)\times U(1)$ 
Higgs model composed of two dual photons ($1^+$) 
(which are dual to two abelian gluons $A^3$ and $A^8$ after 
abelian projection) and scalar ($0^+$) bosons coupled to the dual photons.  
Actually the present author and his collaborators have derived such a model 
called dual Ginzburg-Landau (DGL) model starting from QCD
\cite{suzu88,maedan89,maedan90,monden,kamizawa,matsurev,suzurev}.
After summing up all contributions from closed loops of monopole currents 
appearing after abelian projection, 
the model is composed of two degenerate  dual photons $C^3_{\mu}(x)$ and 
$C^8_{\mu}(x)$ ($1^+$)
and three degenerate complex scalar ($0^+$) fields $\chi_i(x)$ ($i=1\sim 3$) 
with magnetic charges 
as well as quarks and gluons which play the role of simple charged particles.
 Let me call the former dual photons as strong bosons and the latter 
magnetically charged scalar 
as monopole particles. In the confinement phase, strong bosons become massive 
due to the Higgs mechanism and massive monopole particles appear 
in addition to usual hadrons composed of quarks and glueballs.  
When we neglect dynamical quarks and charged gluons for the moment
and consider only an external electric current 
${\vec j}^\beta_{ext}=(j^3_{ext},j^8_{ext})$, 
the model is written as\cite{suzu88,maedan89,monden}
\begin{eqnarray}
  {{\cal L}_{eff}} & = & -\frac 14 {\vec H}_{\mu\nu}^2 
              +\sum_{\alpha=1}^{3} \{ \vert ( \partial_\mu 
                 + ig_m {\vec\epsilon}_\alpha \cdot {\vec C}_\mu )
                   \chi_\alpha \vert ^2 
              - \lambda ( \vert \chi_{\alpha} \vert ^2 
                 - v^2 )^2 \} \nonumber \\
            && \hspace{1cm} - \lambda'( \sum_{\alpha=1}^{3}\vert \chi_\alpha 
                 \vert ^2 - 3 v^2 )^2 
              + \kappa \chi_1 \chi_2 \chi_3 ,
\end{eqnarray}
 where 
\begin{eqnarray*}
&&{\vec \epsilon}_1=( 1 , 0 )\ , {\vec \epsilon}_2=
( -\frac 12 , -\frac {\sqrt{3}}{2} )\ , {\vec \epsilon}_2 = ( -\frac 12,
 \frac {\sqrt{3}}{2} )\ , \hspace{.3cm}
\Im( \chi_1 \chi_2 \chi_3 )= 0, \\
 &&{\vec H}_{\mu\nu} = \partial_\mu {\vec C}_\nu - \partial_\nu {\vec
C}_\mu + \epsilon_{\mu\nu\alpha\beta} n^\alpha (n\cdot \partial)^{-1}
{\vec j}^\beta_{ext}\ , \hspace{.4cm}
{\vec C_\mu} = (C_{\mu}^3, C_{\mu}^8).
\end{eqnarray*} 

In unitary gauge $Im \chi_\alpha =0 $, the classical field equations
\begin{eqnarray}
 \partial_\mu {\vec H}^{\mu\nu} + 2 g_m^2 \sum_{\alpha=1}^{3} 
{\vec \epsilon}_\alpha \cdot ( {\vec \epsilon}_\alpha \cdot {\vec
C}^\nu )
\chi_\alpha^2  =  0  \label{eq1} \\
 \partial_\mu \partial^\mu \chi_\alpha - g_m^2 ( {\vec
\epsilon}_\alpha\cdot {\vec C}_\mu )^2 \chi_\alpha       
   + 2 \lambda ( \chi_\alpha^2 - v^2 )\chi_\alpha =  0. \label{eq2} 
\end{eqnarray}
where $\lambda' = 0$ and $\kappa = 0$ are assumed for simplicity.  In
case of static hadrons we set static charge configurations in ${\vec
j}^\beta_{ext}$.

The model can reproduce analytically the linear potentials between static 
quark-antiquark (meson)\cite{suzu88,maedan89,maedan90,suganuma} 
and also between three quarks (baryon)
\cite{kamizawa}. It can also explain the characteristic 
features of finite-temperature transition of pure QCD found 
by Monte-Carlo simulations, that is, the first (second) order phase transition
in $SU(3)$ $(SU(2))$ QCD\cite{monden}.
A long-range Van der Waals force is shown not to appear between meson-meson 
interactions\cite{matsurev}.
Monopole condensation seems to enhance chiral symmetry breaking\cite{suganuma,matsurev}.

Both strong bosons and monopole particles are neutral with respect to electric 
$U(1)\times U(1)$ and so are proved to be 
physical which are composed of gluons (glueball-like states).
To search for such bosons and to establish them experimentally are therefore 
very crutial in order to prove the correctness of the dual Meissner effect.
Also, numerical Monte-Carlo measurements of such particles 
on large enough lattices in the framework of lattice QCD is 
very important in order to prepare for real experiments. 

\section{Estimate of the masses of new bosons}
The phenomenological analyses of the DGL model lead us to predict existence 
of strong bosons and monopole particles having masses of the order 
O(1GeV), that is , 0.5GeV$\sim$2.0GeV\cite{maedan90,suganuma}
. The value of the mass can not be fixed 
definitely  at present, but it can not be too 
large, because they are related to the value of the string tension 
$\sqrt{\sigma}\sim 450$MeV and the QCD gauge coupling constant $g$ through 
Dirac's quantization condition $gg_m=4\pi n$ ($n=$ integer but $n=1$ is 
actually considered). 

Introducing a static quark and antiquark source 
\begin{eqnarray}
  {\vec j}^\mu_{ext} = {\vec Q} g^{\mu 0} \delta(x) \delta(y) \{ 
         \delta(z-\frac R2) - \delta(z+\frac R2) \},
\end{eqnarray}
where ${\vec Q} = (g/2, g/2\sqrt3)$ and 
$n^\mu = ( 0, 0, 0, 1)$, we can evaluate the string tension $\sigma$ 
by numerically solving the equations of motions (\ref{eq1}) and (\ref{eq2})
\cite{suzu88,maedan89,maedan90,suganuma}.

Especially, exact results can be derived analytically in the extreme type 2 
case (where the Ginzburg-Landau parameter 
$\kappa=\sqrt{2\lambda}/(\sqrt{3}g_m)\gg 1/\sqrt{2}$) and also 
at the border between 
type 1 and type 2 ($\kappa = 1/\sqrt{2}$) as is well known in the 
usual superconductor case.

In the extreme type 2 case, I (partially with Maedan) derived 
\begin{eqnarray}
\sigma=\frac{\vec{Q}^2 m_c^2}{4\pi}
K_0\left(\frac{\sqrt{2}m_c}{m_{\chi}}\right), 
\end{eqnarray}
where a natural infrared cutoff is introduced and 
$K_0$ is a modified Bessel function.
Recently, Suganuma et al.\cite{suganuma} have pointed out that in this case 
we need not introduce the infrared cutoff and have obtained 
\begin{eqnarray}
\sigma=\frac{\vec{Q}^2 m_c^2}{8\pi}
\ln\left(\frac{m_c^2+m_{\chi}^2}{m_c^2}\right), 
\end{eqnarray}
In the extreme type 2 case $m_{\chi}\gg m_c$, 
both give about the same results 
\begin{eqnarray}
\sigma=4\pi v^2\ln\left(\frac{m_{\chi}}{m_c}\right),
\end{eqnarray}
where we have used $<\chi_1>=<\chi_2>=<\chi_3>=v$, $gg_m=4\pi$ and 
$m_c=\sqrt{3}g_m v$\cite{maedan89}.
Adopting $\sigma=(450)^2$ MeV$^2$ from the Cornell potential fit to charmonium
spectra, we get, say, for $m_{\chi}/m_c=1.5\sim 4$,
\begin{eqnarray}
v=198 \sim 108 (\rm{MeV}).
\end{eqnarray}

On the other hand, at the border between type 1 and type 2 cases, 
one can get the first integral of the equations of motions 
(\ref{eq1}-\ref{eq2}):
\begin{eqnarray}
\sqrt{\frac{2}{3}}\frac{1}{\rho}
\frac{\partial}{\partial\rho}\left(\rho\tilde{C}\right)
=\sqrt{2\lambda}(v^2-\chi^2), \label{fint}
\end{eqnarray}
where it is enough to consider only one common 
$\chi$  and  $C$ fields\cite{maedan90}. 
Also $C=C_D + \tilde{C}$ where 
$C_D$ is the Coulomb part and the cylindrical coordinate $(\rho,\theta,z)$ is 
adopted.
When two sources are far apart, the string tension is expressed by 
\begin{eqnarray}
\sigma=|\int\rho d\rho d\theta [2\lambda(v^4-\chi^4)
-2g_m^2\chi^2(C_D+\tilde{C})\tilde{C}]|,
\end{eqnarray}
which reduces using (\ref{fint}) to 
\begin{eqnarray}
\sigma&=&2\pi|\int_0^{\infty} d\rho \frac{2\sqrt{\lambda}}{\sqrt{3}}
\frac{\partial}{\partial\rho}[(\rho\tilde{C})(v^2+\chi^2)]|\\
&=& 4\pi v^2,
\end{eqnarray}
where we have used $\tilde{C}\rightarrow -C_D\rightarrow -g/(4\pi\rho)$ and
$\chi\rightarrow v$ 
 as $\rho\rightarrow \infty$.
Hence we get in this case
\begin{eqnarray}
v=127 (\rm{MeV}).
\end{eqnarray}

The mass $m_c$ depends on the value $g$. Non-perturbative effects 
are not known and  $g=2\sim 5$ may be possible\cite{suganuma}.
Then we get 
\begin{eqnarray}
m_c&=& 0.5\sim 2.0 (\rm{GeV}) \rm{\ \ for\  the\  extreme\  type\ 2\ case},\\
m_c&=& m_{\chi} = 0.6\sim 1.4 (\rm{GeV}) \rm{\ \ for\ the\ border\ case}.
\end{eqnarray}

The masses can be determined also from the abelian electric flux distribution
 and the correlation between the electric flux and the rotation of monopole 
currents as done similarly in the superconductor. 
The Monte-Carlo measurements have been done by some groups
\cite{hay,cea,matsu94}. Although the lattices used are not large enough,
the $SU(3)$ data suggest both masses are almost equal and 
of order $1.5\sim 2.0$ GeV.
Namely the QCD vacuum seems near  the border between type 1 and type 2 
magnetic superconductor. This is consistent with numerical analyses of 
the DGL model\cite{maedan90} and a preliminary Monte-Carlo evaluation of 
axial-vector and scalar correlations using abelian Wilson loops\cite{tezuka}.

Considering the vacuum seems near the border between type 1 and type 2 
magnetic superconductor, we could guess both masses are between 
0.5 GeV and 2.0 GeV. In this rough sense, the new bosons are predicted to 
have the mass of O(1GeV).

\section{Selection rules from the Weyl smmetry}
The above Weyl symmetry is expected to lead us to interesting selection rules 
with respect to transition matrix elements of the strong bosons  
and the monopole particles.

\subsection{Weyl transformation properties of new boson operators}
Ordinary color singlets mesons $\sum_i\bar{\psi}^i\psi^i$ and 
baryons $\sum_{i,j,k=1}^{3}\epsilon_{ijk}\psi_{f1}^i\psi_{f2}^j\psi_{f3}^k$ 
are naturally invariant under the Weyl group. 
Other $U(1)\times U(1)$ neutral hadrons composed of quarks and gluons such as 
$\bar{\psi}\lambda_3 \psi$ are 
Weyl nontrivial. 

What about the new bosons? The strong bosons $C^3$ and $C^8$ have the same 
transformation property as those of the abelian gauge fields $A^3$ and $A^8$
after abelian projection, since they are canonical conjugates and are not 
independent. Hence the strong bosons are 
Weyl nontrivial. The Weyl symmetry is common in the original and in the dual 
expressions of the abelian projected QCD.
However, it is easy to prove that the followings are Weyl invariant:
\begin{eqnarray}
&&(C^3)^2+(C^8)^2 = C^{+}C^{-},\\
&&\bar{\psi}(C^3\lambda_3+C^8\lambda_8)\psi
=\bar{\psi}(C^{+}\lambda_{-}+C^{-}\lambda_{+})\psi,
\end{eqnarray}
where $C^{\pm}\equiv (C^3\pm iC^8)/\sqrt{2}$ and 
$\lambda_{\pm}\equiv (\lambda_3\pm i\lambda_8)/\sqrt{2}$.  
Here we have not written the space-time dependence explicitly.

The monopole fields $\chi_{\alpha}$ have the coupling with 
the strong bosons as follows:
\begin{eqnarray}
          \sum_{\alpha=1}^{3} \{ \vert ( \partial_\mu 
                 + ig_m {\vec\epsilon}_\alpha \cdot {\vec C}_\mu )
                   \chi_\alpha \vert ^2. 
\end{eqnarray}
Hence each $\chi_{\alpha}$ is Weyl nontrivial. Actually,
it is easy to see the strong boson triplet 
$\vec{\epsilon}_{\alpha}\cdot\vec{C}$ ($\alpha=1\sim 3$) and the monopole 
triplet $\chi_{\alpha}$ ($\alpha=1\sim 3$) changes  each other under 
any Weyl transformation. For example, under the permutation (31),
\begin{eqnarray}
\frac{-C^3-\sqrt{3}C^8}{2} &\rightarrow& \frac{C^3+\sqrt{3}C^8}{2},\\
\frac{-C^3+\sqrt{3}C^8}{2} &\rightarrow& -C^3,\\
\chi_2 \rightarrow \chi_2^*\ , &&\hspace{.5cm} \chi_3 \rightarrow \chi_1^*\ .
\end{eqnarray}
Also under the cyclic permutation (123), 
\begin{eqnarray}
\frac{-C^3-\sqrt{3}C^8}{2} &\rightarrow& C^3, \hspace{.5cm}
C^3 \rightarrow \frac{-C^3+\sqrt{3}C^8}{2},\\
\frac{-C^3+\sqrt{3}C^8}{2} &\rightarrow& \frac{-C^3-\sqrt{3}C^8}{2},\\
\chi_1 \rightarrow \chi_2\ , \hspace{.5cm} \chi_2 &\rightarrow& \chi_3\ ,
\hspace{.5cm} \chi_3 \rightarrow \chi_1\ .
\end{eqnarray}

However the mixed state 
\begin{eqnarray}
\chi^0\equiv |\chi_1+\chi_2+\chi_3|/\sqrt{3} 
\end{eqnarray}
is Weyl trivial.

\subsection{The Weyl property of the vacuum}
To fix the transformation properties of the new states, one has to study 
the vacuum. Does the vacuum respect the Weyl symmetry?

In the framework of the DGL theory, the vacuum can be fixed by the 
self-interaction terms of monopole fields:
\begin{eqnarray}
         V= \lambda (\sum_{\alpha=1}^3 \vert \chi_{\alpha} 
           \vert ^2)^2  
             + \lambda'( \sum_{\alpha=1}^{3}\vert \chi_\alpha 
                 \vert ^2 )^2  + \kappa \chi_1 \chi_2 \chi_3 
-\mu^2\sum_{\alpha=1}^3 
           \vert \chi_{\alpha} \vert ^2.
\end{eqnarray}
When magnetic $U(1)\times U(1)$ is spontaneously broken ($\mu^2>0$), 
both vacuum states with spontaneous broken and unbroken 
Weyl symmetry are possible, 
depending on the choice of the parameters $\lambda, \lambda'$ and $\kappa$.
On the other hand, only the symmetric vacuum is chosen in the case of 
$SU(2)$ vacuum.

However, MC simulations of abelian projection of lattice QCD strongly 
suggest that QCD vacuum also respects the Weyl symmetry.
After abelian projection in lattice $SU(3)$ QCD, there are two 
independent abelian link variables corresponding to $A^3$ and $A^8$.
The value of the string tension, the Polyakov loops and the masses of the 
strong bosons are seen to be the same when we evaluate them in terms of each 
abelian link variable, although the fact is not explicitly written in the 
published papers\cite{matsu94,suzu95a,suzu95c}.
This suggests that the $SU(3)$ QCD vacuum respects the Weyl symmetry.

In the previous section, we have taken $\lambda'=0$ 
in which the Weyl symmetry is not broken spontaneously.
Hence we have tacitly assumed the invariance of the vacuum as suggested 
in the MC simulation.

\subsection{Selection rules}
In the folowing also, we assume that the Weyl symmetry is not spontaneously 
broken in the $SU(3)$ QCD vacuum.
Since the strong bosons have the mass of O(1GeV), it is natural 
to suppose that they are 
the lightest Weyl nontrivial states which can couple directly to quarks and 
gluons.
Considering that the QCD Hamiltonian 
is invariant under the Weyl symmetry, 
we can prove that any matrix element between $C_{\mu}^3\ (C_{\mu}^8)$
and ordinary hadron states denoted by $|h\rangle$ vanishes. 
Applying the $(12)$ permutation, we get 
\begin{equation}
\langle C_{\mu}^3 |H|h\rangle = -\langle C_{\mu}^3 |H|h\rangle =0.
\end{equation}
Also under the (31) permutation, we have 
\begin{equation}
\langle C_{\mu}^3+\sqrt{3}C_{\mu}^8 |H|h\rangle 
= -\langle C_{\mu}^3+\sqrt{3}C_{\mu}^8 |H|h\rangle =0.
\end{equation}
Hence 
\begin{equation}
\langle C_{\mu}^8 |H|h\rangle = 0.
\end{equation}

The Weyl trivial states can couple to ordinary hadrons. 
Hence such a state as $C^{+}C^{-}$ can decay into or produced by ordinary 
color singlet hadrons. Also Weyl trivial $\chi^0$ 
 can couple to ordinary hadrons through 
$C^+C^-$.

Now one can understand why such a light axial vector state has not been found 
in the usual lattice search of glueballs. Usually, Wilson loops composed of 
a full $SU(3)$ link field are used to search for glueball states. 
But such Wilson loops correspond to totally color singlet Weyl trivial states.
Hence one can not get any information of such Weyl nontrivial states like 
the strong boson. On the otherhand, abelian Wilson loops composed 
of abelian link fields alone after abelian projection 
are in general  Weyl nontrivial. Actually, 
Monte-Carlo simulations using such abelian Wilson loops give the mass 
of O(1GeV)\cite{hay,cea,matsu94}, although the 
lattice size is not large enough.

\section{Production and annihilation of new bosons and experiments}
Is it possible to evaluate matrix elements of (pair) production 
or pair annihilation of 
the new bosons  analytically? 
It is very interesting and challenging, but there are some severe problems:
\begin{enumerate}
\item
If we introduce dynamical charged quarks into the DGL model
(still neglecting dynamical charged gluons), we get the following 
Lagrangian:
\begin{eqnarray}
  {{\cal L}_{eff}} & = & -\frac 14 {\vec H}_{\mu\nu}^2 
              +\sum_{\alpha=1}^{3} \{ \vert ( \partial_\mu 
                 + ig_m {\vec\epsilon}_\alpha \cdot {\vec C}_\mu )
                   \chi_\alpha \vert ^2 
              - \lambda ( \vert \chi_{\alpha} \vert ^2 
                 - v^2 )^2 \} \nonumber \\
            && \hspace{1cm} + \lambda'( \sum_{\alpha=1}^{3}\vert \chi_\alpha 
                 \vert ^2 - 3 v^2 )^2 
              + \kappa \chi_1 \chi_2 \chi_3 
             +\bar{\psi}(i\partial\!\!\!\!{/}-m)\psi\ ,\label{newlag}
\end{eqnarray}
where
\begin{eqnarray*}
{\vec H}_{\mu\nu} &=& \partial_\mu {\vec C}_\nu - \partial_\nu {\vec
C}_\mu + \epsilon_{\mu\nu\alpha\beta} n^\alpha (n\cdot \partial)^{-1}
{\vec j}^\beta\ , \\
{\vec j}^\mu(x)&=&-g\,\bar{\psi}(x)\gamma^\mu \frac {{\vec \lambda}}2
 \psi (x)\ ,\ \ {\vec \lambda}=(\lambda^3,\lambda^8)\ .\\
\end{eqnarray*}
Since the theory contains two coupling constants $g$ and $g_m$ satisfying 
the Dirac quantization condition $gg_m=4\pi$, a perturbative treatment 
is impossible. We have to resort to some nonperturbative method.
\item
Moreover, in the DGL model, electrically charged quarks and gluons are 
topological quantities just as monopoles in the original QCD. There must arise 
inevitablly non-local interactions 
between dynamical and topological quantities. This reflects the necessity of 
the Dirac string and actually there are non-local terms containing 
$n_\lambda(n\cdot\partial)^{-1}$ in (\ref{newlag}).
\end{enumerate}

Maybe, the most reliable method is Monte-Carlo simulations of lattice QCD.
The new boson state with non-trivial Weyl property 
can be constructed in terms of abelian Wilson loops after abelian projection.
If we evaluate correlations of such operators and ordinary hadron operators 
composed of full Wilson loops, we would get information 
of the matrix elements of, say, pair annihilation of the 
strong bosons into ordinary hadrons, although we need large lattices and 
very long CPU time. 
But this is worth while to be challenged.

Experimentally, there may be  severe constraints with repect to such matrix 
elements\cite{berkeley}.  
They could be used to test the correctness of the 'tHooft mechanism.
Here I only list up some possible examples:
\begin{itemize}
\item
$e^+ + e^- \rightarrow \gamma$ + $X^0$, where $X^0$ is a pair of 
the strong bosons or $\chi^0$.  
\item
$J/\psi$ (and $\Upsilon$) $\rightarrow \gamma$ + $X^0$.
In this case, the $\gamma + \chi^0$ decay seems severely restricted.
\item
$\bar{p} + p \rightarrow$ $C^+ + C^-$.
\item
$\pi^- + p \rightarrow n$ + $X^0$.
\end{itemize}

If the couplings of $C^+ + C^- \rightarrow$ ordinary hadrons 
happen to be unexpectedly small due to some unknown mechanism, 
the new bosons might be a new candidate 
of the dark matter. Such new bosons are produced much through the 
transition from quark-gluon phase to hadron phase. 

\vspace{1cm}
I would like to express my thanks to Y.Matsubara for various 
useful discussions and H.Shiba for pointing  careless 
mistakes in the matrix representation of the Weyl transformation and then 
in the transformation property of ordinary baryons. K.Aoki is deeply 
acknowledged for discussions clarifying the property of the Weyl symmetry.
This work is financially supported by JSPS Grant-in Aid for Scientific  
Research (B) (No.06452028).


\begin{thebibliography}{99}
\bibitem{thooft2} G. 'tHooft, Nucl. Phys. B190 (1981) 455.
\bibitem{suzu90} T. Suzuki and I. Yotsuyanagi, Phys. Rev. D42 (1990) 4257;
\ Nucl.Phys.B(Proc.Suppl.)20 (1991) 236.
\bibitem{hio91b} S. Hioki et al., Phys. Lett. 272B (1991) 326; errata, 
Phys. Lett. B281 (1992) 416;\  Nucl.Phys. B(Proc.Suppl.) 26 (1992) 441.
\bibitem{suzu93} T. Suzuki, Nucl. Phys. B(Proc. Suppl.) 30 (1993) 176.  
\bibitem{suzu95a} T. Suzuki et al., Phys. Lett. B347 (1995) 375;
\ Nucl. Phys. B(Proc. Suppl.) 42 (1995) 529.  
\bibitem{miya95a} O.Miyamura, Phys.Lett. B353 (1995) 91. 
\bibitem{miya95b} O.Miyamura and S.Origuchi, 
Hiroshima Univ. Report hep-lat 9508015.
\bibitem{suzu95c} T. Suzuki et al., Talk at 'Lattice 95'. To appear in 
\ Nucl. Phys. B(Proc. Suppl.).  hep-lat 9509016.
\bibitem{ejiri95b} S. Ejiri et al., Talk at 'Lattice 95'. To appear in 
\ Nucl. Phys. B(Proc. Suppl.). hep-lat 9509013.
\bibitem{suzurev} T. Suzuki, See the reviews in  conferences and workshops.
Int. School-Seminar '93 - Hadrons and Nuclei from QCD - 
(World Scientific, 1994)   325; \ YITP Workshop "From Hadronic Matter 
to Quark Matter" to appear in Prog.
Theor. Phys. Suppl.; \ German-Japan Seminar on Massively Parallel Computers 
(World Scientific, 1995); 
\ RCNP Workshop on Color Confinement and Hadrons (Osaka 1995)
'Confinement 95' to appear in Prog. Theor. Phys. Suppl.. ; \ 
ECT Workshop 'Nonperturbative Approaches to QCD' (Trento 1995).
\bibitem{kron} A.S. Kronfeld et al., Phys. Lett. 198B (1987) 516, \\
               A.S. Kronfeld et al., Nucl.Phys. B293 (1987) 461.
\bibitem{hio91a} S. Hioki et al., Phys. Lett. 271B (1991) 201. 
\bibitem{shier} F. Brandstaeter et al., Phys. Lett. 272B (1991) 319.
\bibitem{kita93} S. Kitahara et al., Nucl Phys. B(proc. Suppl.) 30 (1993) 557.
\bibitem{stack} J.D. Stack and R.J. Wensley, Nucl.Phys. B371 (1992) 597;
Talk at Lattice 95. To appear in 
\ Nucl. Phys. B(Proc. Suppl.). 
\bibitem{shiba94b} H. Shiba and T. Suzuki, Phys. Lett. B333 (1994) 461.
\bibitem{ejiri95a} S.Ejiri et al., Phys. Lett. B343 (1995) 315; Nucl. Phys. 
B(Proc. Suppl.) 42 (1995) 481.
\bibitem{ejiri95c} S. Ejiri, Talk at 'Lattice 95'. To appear in 
\ Nucl. Phys. B(Proc. Suppl.). hep-lat 9509014.
\bibitem{born} V.G. Bornyakov et al., Phys. Lett. 284B (1992) 99.
\bibitem{hio92} S. Hioki et al., Phys. Lett. 285B (1992) 100; 
 Nucl.Phys. B(Proc.Suppl.) 26 (1992) 450.
\bibitem{ivanenko} T.L. Ivanenko et al., Phys. Lett. 252B (1990) 631.
\bibitem{shiba94a} H. Shiba and T. Suzuki, 
 Nucl.Phys. B(Proc.Suppl.) 34 (1994) 182; 
 Nucl.Phys. B(Proc.Suppl.) 42 (1995) 282; 
Phys. Lett. B351 (1995) 519.
\bibitem{suzu95b} T. Suzuki et al., Talk at 'Lattice 95'. To appear in 
\ Nucl. Phys. B(Proc. Suppl.).  hep-lat 9509015.
\bibitem{kita95} S. Kitahara et al., Prog. Theor. Phys. 93 (1995) 1; 
 Nucl.Phys. B(Proc.Suppl.) 42 (1995) 511.
\bibitem{hay} V. Singh et al., LSU preprint LSUHEP- 1- 92 (1992);
 Nucl. Phys. B(Proc. Suppl.) 30 (1993) 568.  
\bibitem{cea} P. Cea and L. Cosmai, 
 Nucl. Phys. B(Proc. Suppl.) 30 (1993) 572.  
\bibitem{matsu94} Y. Matsubara et al., 
 Nucl. Phys. B(Proc. Suppl.) 34 (1994) 176.  
\bibitem{giacomo} L. Del Debbio et al., To appear in Phys. Lett.B; 
 Talk at 'Lattice 95'. To appear in \ Nucl. Phys. B(Proc. Suppl.).  
\bibitem{arafune} J. Arafune $et$ $al.$, Jour. Math. Phys. {\bf 16}
 (1975) 433.
\bibitem{suzu88} T. Suzuki, Prog. Theor. Phys. 81 (1988) 929; 81 (1989) 752.
\bibitem{maedan89} S. Maedan and T. Suzuki, Prog. Theor. Phys. 81 (1989) 229.
\bibitem{maedan90} S. Maedan et al., Prog. Theor. Phys. 84 (1990) 130.
\bibitem{monden}  
H. Monden et al., Phys. Lett. B294 (1992) 100.
\bibitem{kamizawa} S. Kamizawa et al., Nucl Phys. B389 (1993) 563.
\bibitem{matsurev} Y. Matsubara, Review talk in 
ECT Workshop 'Nonperturbative Approaches to QCD' (Trento 1995).
\bibitem{suganuma} H. Suganuma et al., Nucl. Phys. B435 (1995) 207; 
Talks in 
 RCNP Workshop on Color Confinement and Hadrons (Osaka 1995); \ 
ECT Workshop 'Nonperturbative Approaches to QCD'(Trento 1995).
\bibitem{tezuka} Y.Tezuka, Master thesis submitted to Kanazawa Univ. 
(unpublished March, 1995).
\bibitem{berkeley} Particle Data Group, Phys. Rev. D50 (1994) 1173 
and references therein.

\end{thebibliography}
\end{document}